\documentclass[12pt,twoside]{article}
\usepackage{fleqn,espcrc1}

\usepackage{graphicx}
\usepackage[figuresright]{rotating}

\newcommand{\AmS}{{\protect\the\textfont2
  A\kern-.1667em\lower.5ex\hbox{M}\kern-.125emS}}

\title{Hyperon effects on the properties of $\beta$-stable neutron star matter}

\author{I. Vida\~na$^{\rm a}$, A. Polls$^{\rm a}$, A. Ramos\address{Departament d'Estructura 
i Constituents de la Mat\`eria, Universitat de Barcelona, E-08028 Barcelona, Spain}, 
L. Engvik$^{\rm b}$ and M. Hjorth-Jensen\address{Department of Physics, University of Oslo, N-0316 
Oslo, Norway}}
       
\begin{document}

\maketitle

\begin{abstract}


We present results from Brueckner--Hartree--Fock calculations for $\beta$-stable neutron star matter with nucleonic
and hyperonic degrees of freedom employing the most recent parametrizations of the baryon-baryon interaction of the 
Nijmegen group. Only $\Sigma^-$ and $\Lambda$ are present up to densities $\sim 7\rho_0$. The corresponding equations of state 
are then used to compute properties of neutron stars such as masses and radii.

\end{abstract}

\section{INTRODUCTION}

Neutron stars offer an interesting interplay between nuclear processes and astrophysical observables. Conditions of matter inside such objects 
are very different from those one can find on Earth, so a good knowledge of the Equation of State (EOS) at such high densities is required to understand the
properties of neutron stars. At densities near to the nuclear saturation density ($\rho_0 \sim 0.16$ fm$^{-3}$), matter is mainly composed of neutrons,
protons and electrons in $\beta$-equilibrium . When the density increases new degrees of freedom may appear such as pion o kaon condenstates, quark matter or
hyperons. The presence of the latter on neutron star matter is the topic of the present work. 

Hyperonic degrees of freedom have been considered by several authors, mainly within the framework of relativistic mean field models
\cite{Pr97} or parametrized effective interactions \cite{Ba97}. Recently Schulze {\it et al.} \cite{Sch98} have performed
many-body calculations with realistic hyperon-nucleon interactions in order to study the onset point of hyperons in neutron star matter and
the effect of nucleonic three-body forces. Nevertheless, they have not considered the role of the hyperon-hyperon interaction, which, however,
is essential as soon as the first hyperon appears in matter, because it modifies the single-particle energies of all the species and, as 
a consequence, the chemical potentials and the equilibrium conditions.

We present results for a microscopic many-body calculation of the Brueckner--Hartree--Fock (BHF) type for $\beta$-stable neutron star matter 
with nucleonic and hyperonic degrees of freedom including not only the nucleon-nucleon and hyperon-nucleon interactions but also the hyperon-hyperon ones. 
Special attention will be paid to the role of this interaction.

The paper is organized as follows. In the next section we will describe very briefly the formalism we have employed. Results for the
composition and the EOS of $\beta$-stable neutron star matter and for the structure of the star will be presented in section 3. A short summary 
and some conclusions will be given in section 4.

\section{FORMALISM}

Our many-body scheme starts with the most recent parametrization of the bare baryon-baryon potential for the complete baryon octet as defined
by Stoks and Rijken in Ref. \cite{St99}. This potential describes all the strangeness sectors from $S=0$ to $S=-4$ and is based on SU(3)
extensions of the Nijmegen nucleon-nucleon and hyperon-nucleon potentials. We introduce effects from the medium by constructing the so-called $G$ matrix
and solving the equations for the single-particle energies of the various baryons self-consistently. The $G$ matrix is formally given by the
solution of the Bethe-Goldstone integral equation
\begin{equation}
G(\omega)=V+V\frac{Q}{\omega-H_0+i\eta}G(\omega) \ .
\label{eq:eq1}
\end{equation}

At lowest order in the BHF theory the single-particle potential $U_{B_i}$ of a baryon $B_i$ is given by
\begin{eqnarray}
U_{B_i}={\mathrm Re} \sum_{B_j\leq F_j}
       \left\langle B_iB_j\right |G(\omega=E_{B_i}+E_{B_j})
       \left | B_i B_j \right\rangle \ , 
\label{eq:eq2}
\end{eqnarray}
and the total non-relativistic energy density, relative to the nucleon mass, reads
\begin{equation}
\epsilon=\epsilon_l+\epsilon_b=\epsilon_l
+2\sum_{B_i}\int_0^{k_F^{(B_i)}}\frac{d^3k}{(2\pi)^3}
\left(M_{B_i}+\frac{\hbar^2k^2}{2M_{B_i}}+\frac{1}{2}U_{B_i}(k)-M_N\right) \ ,
\end{equation}
where $\epsilon_l$ stands for the contribution of the leptons, considered here as a free gas. The chemical potentials are
\begin{equation}
\mu_B=E_B(k_F^{(B)})=M_B+T_B(k_F^{(B)})+U_B^N(k_F^{(B)})+U_B^Y(k_F^{(B)}) \ ,
\end{equation}
where the superscript in the potential denote the interaction of baryon $B$ with nucleons ($N$) or hyperons ($Y$).

In order to reproduce the saturation properties of nuclear matter, we have replaced the pure nucleonic part of the EOS by the variational calculation of
Akmal {\it et al.} \cite{Ak98} which employs the Argonne $V_{18}$ interaction with three-body forces and relativistic boost corrections. More details can be 
found in Refs. \cite{Vi00} and \cite{Vi00bis}.

\section{RESULTS}

\begin{figure}[htb]
\begin{minipage}[t]{75mm}
\includegraphics[width=18pc]{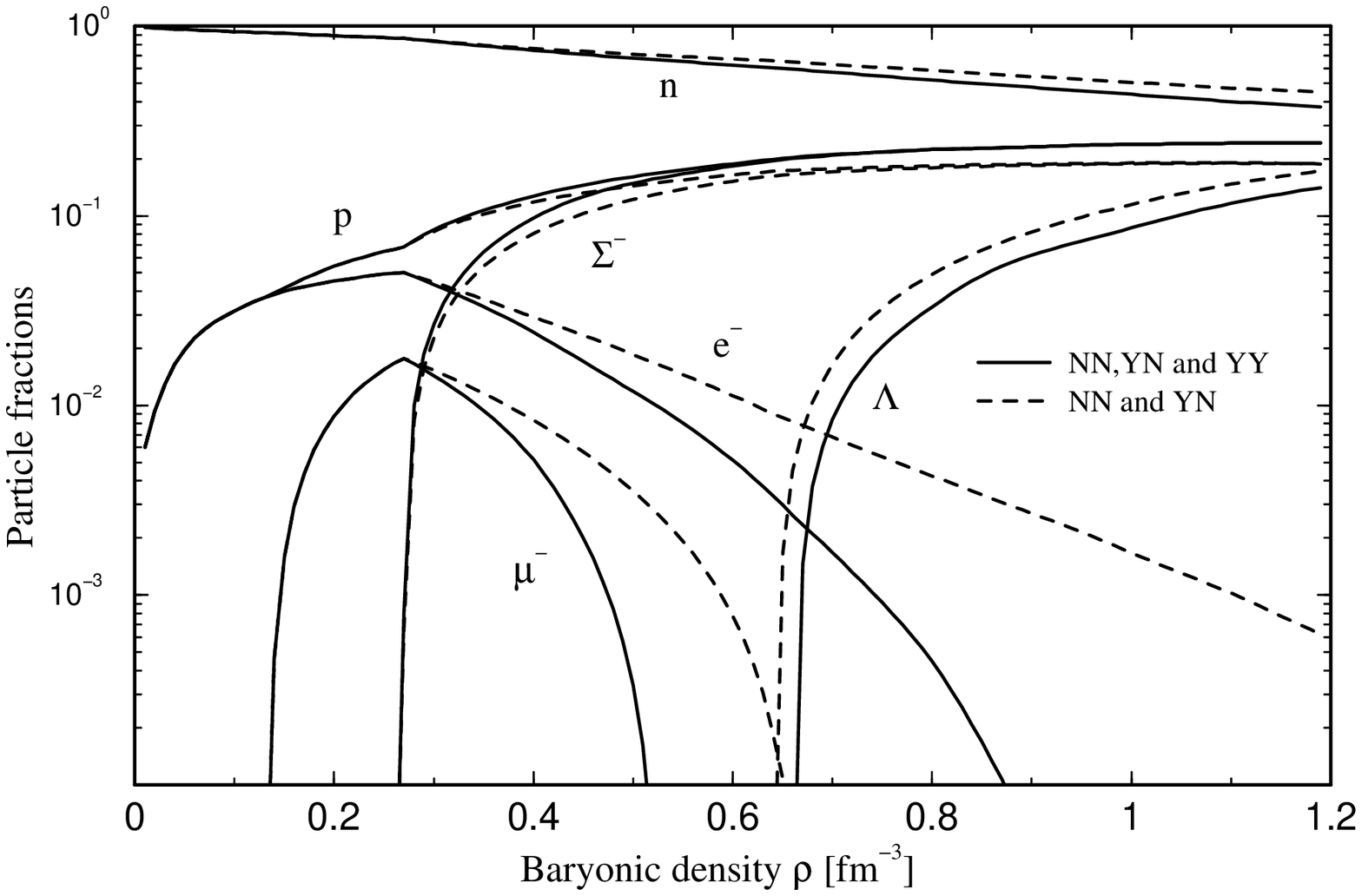}
\caption{Particle fractions as a function of density.}
\label{fig:fig1}
\end{minipage}
\hspace{\fill}
\begin{minipage}[t]{75mm}
\includegraphics[width=18pc]{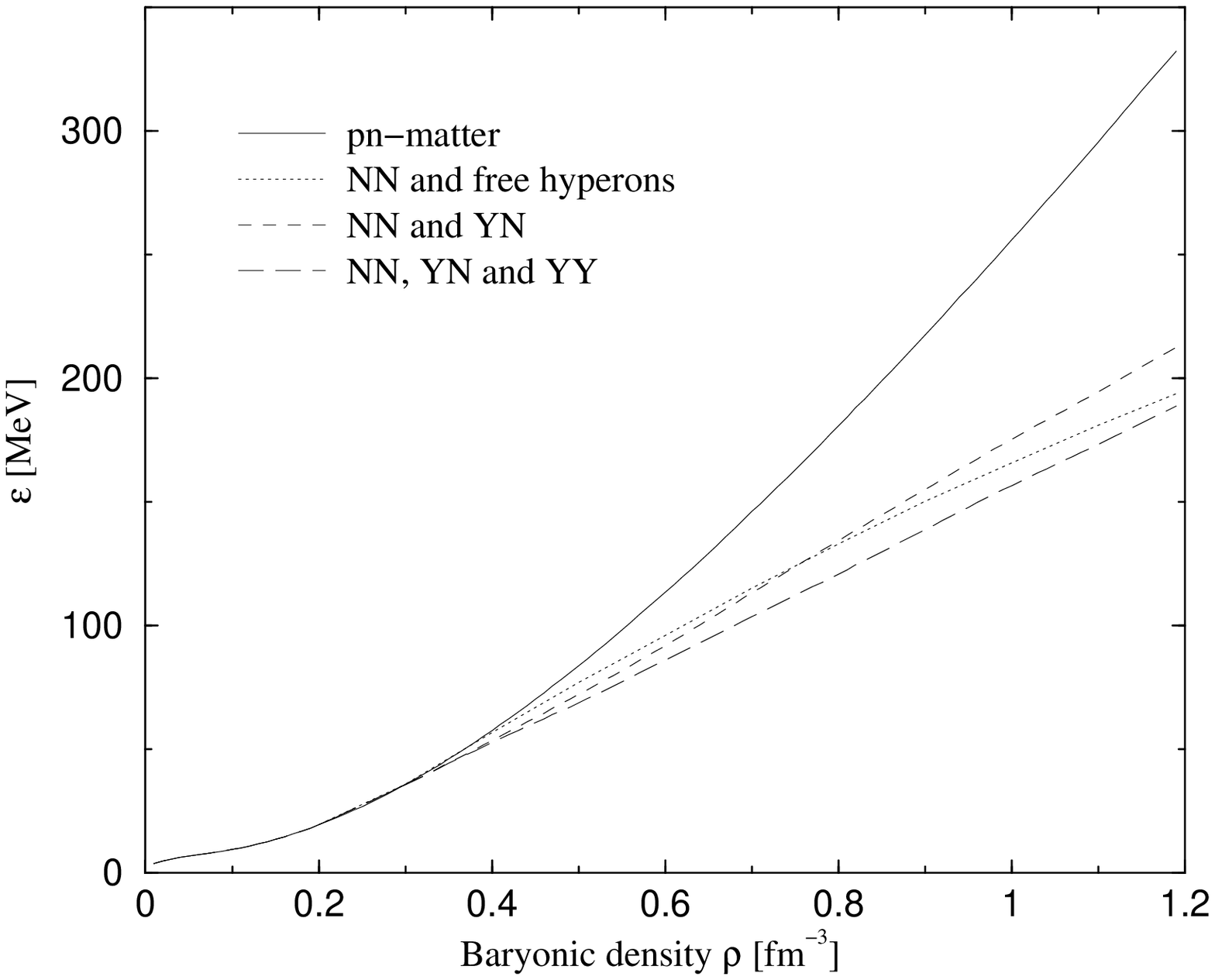}
\caption{Energy per baryon as a function of density.}
\label{fig:fig2}
\end{minipage}
\end{figure}

Fig. \ref{fig:fig1} shows the composition of $\beta$-stable neutron star matter up to density $\rho=1.2$ fm$^{-3}$. Solid lines correspond to a
calculation in which all the interactions ($NN$, $YN$, $YY$) are included, dashed lines show the result when $YY$ interaction is switched off. $\Sigma^-$
appears in both cases at the same density because it is the first hyperon to appear and therefore the $YY$ interaction plays no role for densities below this
point. There is a reduction of its fraction when $YY$ is switched off because of the absence of  the strongly attractive $\Sigma^-\Sigma^-$ interaction. In
turn, a moderate increase on the lepton fraction is observed in order to keep charge neutrality. On the other hand less $\Sigma^-$ implies less $\Sigma^- n$
pairs, whose interaction is very attractive in this model. This means that the neutron chemical potential becomes less attractive and, as a consequence,
$\Lambda$ appears at a smaller density and has a larger relative fraction. 

Fig. \ref{fig:fig2} shows the EOS for four different cases: pure nucleonic matter, matter with nucleons and free hyperons, matter with nucleons and hyperons
interacting only with nucleons, and matter with nucleons and hyperons interacting both with nucleons and hyperons. The appearance of hyperons
leads to a considerable softening of the EOS, which is due essentialy to a reduction of the kinetic energy. The $YN$ interaction has two effects. For
densities up to $\sim 0.72$ fm$^{-3}$, it is attractive and makes the EOS even softer. But for larger densities it is repulsive and the EOS becomes
slightly stiffer. The $YY$ interaction is always attractive producing a softening of the EOS over the whole range of densities explored.

Finally, we have solved the Tolman-Oppenheimer-Volkoff equations, with and without rotational corrections, for three of the EOS discussed in the previous
paragraph. For $\beta$-stable pure nucleonic matter the EOS is rather stiff and yields a maximum mass of $1.89
M_{\odot}$ without rotational corrections and $2.11 M_{\odot}$ when rotational corrections are included, where $M_{\odot}$ is the solar mass. 
A second set of results refers to the case where we
allow for the presence of hyperons
and consider the hyperon-nucleon interaction but explicitly exclude the hyperon-hyperon one. Without rotational correction we obtain a maximum mass 
$1.47 M_{\odot}$, whereas the rotational correction increases the mass to $1.60 M_{\odot}$. This large reduction is mainly a consequence of the strong
softening of the EOS due to the appearance of hyperons. The last EOS includes also the hyperon-hyperon interaction. The inclusion of the
hyperon-hyperon interaction leads to a further softening of the EOS and this leads to an additional reduction of the total mass. In this case whe obtain a maximum
mass of $1.34 M_{\odot}$ when rotational corrections are not included and $1.44 M_{\odot}$ when rotation is taken into account. If other hyperons were to appear
at higher densities, this would most likely lead to a further sosftening of the EOS, and thereby smaller neutron star masses. 

Although we have only considered the formation of hyperons in neutron stars, transitions to other degrees of freedom such as quark matter, kaon condensation, and
pion condensation may or may not take place in neutron star matter. We would, however, like to emphasize that the hyperon formation mechanism is perhaps the most
robust one and is likely to occur in the interior of a neutron star, unless the hyperon self-energies are strongly repulsive due to a possible repulsive
hyperon-nucleon and hyperon-hyperon interactions, a repulsion which would contradict present data on hypernuclei \cite{Ba90}. The EOS with hyperons yields,
however, maximum neutron star masses which are barely compatible with the observations of around $1.4 M_{\odot}$. This means that our EOS with hyperons needs to be stiffer, a
fact which may in turn imply more complicated many-body terms not included in our calculations, such as three-body forces between nucleons and hyperons
\cite{Ya00} and/or relativistic effects, are needed.

\section{SUMMARY AND CONCLUSIONS}

With the most recent parametrization of the bare baryon-baryon potential for the complete octet of baryons, we have performed a microscopic
calculation of the Brueckner--Hartree--Fock type of the structure of $\beta$-stable neutron star matter including nucleonic and hyperonic
degrees of freedom. The potential model employed allows only for the presence of $\Sigma^-$ and $\Lambda$ hyperons up to densities of about
seven times nuclear matter saturation density. 
The presence of hyperons leads to a considerable softening of the EOS, entailing a corresponding reduction of the maximum mass of the neutron
star. The inclusion of the hyperon-hyperon interaction leads to a further softening of the EOS, since this interaction is attractive over the
whole density range explored. This is mainly due to the $\Sigma \Sigma$ interaction which is strongly enough to develop a bound state
\cite{St99}. We note, however, that the $\Lambda \Lambda$ attraction produced by this model is only mild, not being able to reproduce the
experimental $2\Lambda$ separation energy of $\Delta B_{\Lambda \Lambda} \sim 4-5$ MeV. Whether this additional softening is
realistic or not will depend on the details of the hyperon-hyperon interaction that is, unfortunatelly, not well constrained at present. New
data in the $S=-2$ sector, either from double-$\Lambda$ hypernuclei or from $\Xi^-$-atoms, are very much awaited for.

\end{document}